\documentclass[aps,prc,letterpaper,11pt
,twoside,tightenlines,nofootinbib,showpacs
,preprint]{revtex4}
\usepackage{graphicx}
\usepackage[sort&compress]{natbib}
\usepackage{subfigure}
\usepackage{amsmath}
\usepackage{amssymb}
\usepackage{amsfonts}
\usepackage{cancel}
\usepackage{color}
\usepackage{ulem}
\begin{document}

\arraycolsep1.5pt

\newcommand{\Ima}{\textrm{Im}}
\newcommand{\Rea}{\textrm{Re}}
\newcommand{\mev}{\textrm{ MeV}}
\newcommand{\be}{\begin{equation}}
\newcommand{\ee}{\end{equation}}
\newcommand{\ba}{\begin{eqnarray}}
\newcommand{\ea}{\end{eqnarray}}
\newcommand{\gev}{\textrm{ GeV}}
\newcommand{\nn}{{\nonumber}}
\newcommand{\dtres}{d^{\hspace{0.1mm} 3}\hspace{-0.5mm}}

\newcommand{\red}[1]{\textcolor[named]{Red}{#1}}
\newcommand{\blue}[1]{\textcolor[named]{Blue}{#1}}

\def\del{\partial}

\title{$\eta \prime$ nucleus optical potential and possible $\eta
\prime$ bound states}

\author{H.~Nagahiro, S.~Hirenzaki}
\affiliation{
Department of Physics, Nara Women's University, Nara 630-8506, Japan}

\author{E.~Oset}
\affiliation{
Departamento de F\'{\i}sica Te\'orica and IFIC, Centro Mixto Universidad
de Valencia-CSIC,
Institutos de Investigaci\'on de Paterna, Aptdo. 22085, 46071 Valencia,
Spain}

\author{A.~Ramos}
\affiliation{Departament d'Estructura i Constituents de la Mat\`eria and
Institut de Ci\`encies del Cosmos. \\ Universitat de Barcelona,
Avda. Diagonal 645, 08028 Barcelona, Spain}

\date{\today}

 \begin{abstract}

Starting from a recent model of the $\eta^\prime N$ interaction, we
  evaluate the $\eta^\prime$-nucleus optical potential, including the
  contribution of lowest order in density, $t\rho/2 m_{\eta^\prime}$,
  together with the second order terms accounting for $\eta^\prime$
  absorption by two nucleons. We also calculate the formation cross
  section of the $\eta^\prime$ bound states from $(\pi^+,p)$ reactions
  on nuclei. The $\eta^\prime$-nucleus potential suffers from
  uncertainties tied to the poorly known $\eta^\prime N$ interaction,
  which can be partially constrained by the experimental modulus of the
  $\eta^\prime N$ scattering length and/or  the recently measured
  transparency ratios in $\eta^\prime$ nuclear photoproduction. Assuming
  an attractive interaction and taking the claimed experimental value
  $|a_{\eta^\prime N}|=0.1$ fm, we obtain a $\eta^\prime$ optical
  potential in nuclear matter at saturation density of
  $V_{\eta^\prime}=-(8.7+1.8 i)$ MeV, not attractive enough to produce
  $\eta^\prime$ bound states in {light} nuclei. Larger values of the scattering
  length give rise to deeper optical potentials, with moderate enough
  imaginary parts.  For a value $|a_{\eta^\prime N}|=0.3$~fm, which can
  still be considered to lie within the uncertainties of the
  experimental constraints,  the spectra of light and medium nuclei show
  clear structures associated to $\eta^\prime$-nuclear bound states and
  to threshold enhancements in the unbound region.

\end{abstract}
\maketitle

\section{Introduction}
\label{Intro}

The $\eta^\prime$ meson is an important particle to understand QCD dynamics since
it is linked to the $U_A (1) $ axial vector anomaly
\cite{kogut,weinberg,tooft,witten,Kawarabayashi}. Yet, the $\eta^\prime ~ N$
interaction is poorly known. Experimentally, one has estimates of the
modulus of the scattering length, $|a_{\eta^\prime N}|\sim 0.1$ fm
\cite{Moskal:2000pu}. The recent studies of the transparency ratio in
$\eta^\prime$ photoproduction from nuclei indicate that the width $\Gamma_{\eta^\prime}$ of the
$\eta^\prime$ in the nucleus for momenta of the $\eta^\prime $
around 1050 MeV is about 20--30 MeV at nuclear saturation density, $\rho = \rho_0$\cite{nanova}.
Yet, the stability of the transparency ratio as a function of the
$\eta^\prime$ momentum, suggesting a relatively constant inelastic $\eta^\prime N$
cross section, also indicates that $\Gamma_{\eta^\prime}$ ($\simeq v_{\eta^\prime} \sigma \rho$)  should be
smaller at lower energies and values around 10 MeV could be a reasonable
estimate.
%%
%around
%1500 MeV is of about 20-30 MeV at $\rho=\rho_0$,
%but at low
%$\eta^\prime$ momenta it can be of the order of 10 MeV \cite{nanova}.
%%%
On
the other hand we are only aware of one theoretical calculation of the
$\eta^\prime N$ interaction \cite{etaprime}. This calculation is done
within the chiral unitary approach with coupled channels, including the
Weinberg-Tomozawa interaction which acts only on the octet of
pseudoscalar mesons. This requires the use of the mixing angle of the
octet and singlet to give the physical $\eta$ and  $\eta^\prime$ mesons, that is taken from Ref.~\cite{ambrosino}. In addition, one also uses the Lagrangian that
couples the singlet meson to the baryons developed in
Ref.~\cite{borasinglet}. The unknown strength of the singlet Lagrangian
can be
constrained from the value of the modulus of the scattering length, $|a_{\eta^\prime N}|$, but then the sign
of the scattering amplitude remains unknown. Yet, the study of Ref.~\cite{etaprime} finds that the
inelastic cross sections are rather independent on this parameter and
the model makes clear predictions and sets bounds on the scattering
amplitude.

The possible existence of $\eta^\prime$ bound states in nuclei has been
investigated in the past
\cite{Nagahiro:2004qz,Nagahiro:2006dr,Jido:2011pq} where the mass
of the $\eta^\prime$ in the medium was estimated from arguments of chiral symmetry
restoration. A reduction of the $\eta^\prime$ mass in the medium of
about 100 MeV was obtained  in  \cite{Jido:2011pq} and, invoking a small
imaginary part of the optical potential, relatively narrow bound states
were predicted and reactions to eventually measure them were also suggested.
The modification of the
$\eta^\prime$ mass has also been studied using various effective
lagrangians of the Nambu-Jona-Lasinio type~\cite{Bernard:1987sx,Kunihiro:1989my,Costa:2002gk}
or within the quark-meson coupling model \cite{tsushima}. None of these
works included an imaginary component of the $\eta^\prime$ optical
potential accounting for the effect of inelastic channels or
multinucleon absorption. 

In the present work we use the model of  Ref.~\cite{etaprime} to derive
the $\eta^\prime$ nucleus optical potential and investigate the
possibility of obtaining $\eta^\prime$ bound states in nuclei. 
As discussed in Ref.~\cite{etaprime},
the coupling of the singlet meson to the baryons
contributes mostly to the $\eta^\prime$ elastic channel and barely modifies
the inelastic channels.  This unique
feature resembles that of the anomaly effects suggested in
Ref.~\cite{Jido:2011pq} using arguments of symmetry.  Since the
value of the singlet coupling is a free parameter of the theory, we can
vary its strength up to values that simulate the scenario proposed in
Ref.~\cite{Jido:2011pq}. 

In addition to the optical potential obtained from the selfenergy at
lowest order in the nuclear density, $t \rho$,  we also calculate, using
standard many body techniques, the second order density contributions
which account for $\eta^\prime$ absorption by pairs of nucleons. 
We study several options and use the
experimental constraints  \cite{Moskal:2000pu,nanova} to obtain likely
optical potentials with which we 
evaluate the bound states in different nuclei. Assuming the potential to
be attractive, we find bound states of $\eta^\prime$ in most nuclei,
with their corresponding widths, which should be
taken into consideration in an eventual experimental search.  With this study we aim at getting a
deeper insight and new information on the properties of the
$\eta^\prime(958)$ meson in nuclei and on possible effects of the
$U_A(1)$ anomaly with the partial restoration of chiral symmetry at
finite density. 

\section{Formalism}
The $\eta^\prime$ selfenergy  in nuclear
matter at lowest order in the density is given by
\be
\Pi^{\rm 1st}= t\rho\ ,
\ee
where $t$ is the particle-nucleon scattering matrix. The optical
potential is related to it by 

\be
V_{\rm opt}^{\rm 1st}=\frac{ \Pi^{\rm 1st} }{ 2 \omega_{\eta^\prime}}\ ,
\ee
where $\omega_{\eta^\prime}$ is the energy of $\eta^\prime$.

In the present work, we evaluate the $\eta^\prime$ optical potential for a
$\eta^\prime$ meson at rest ($\vec{p}_{\eta^\prime}\sim 0$). We use the $\eta^\prime N$ scattering amplitudes of \cite{etaprime}
for different values of $a_{\eta^\prime N}$. Since we are only
interested in the case where the interaction is attractive, we take the
values of $t$ corresponding to Re($a_{\eta^\prime N}$) positive.  In Table \ref{table:amplitude} we
show the values of $t$ for the
$\eta^\prime N \to \eta^\prime N$ elastic amplitude and the
$\eta^\prime N \to \eta N$ transition amplitude, which will play
some role in the absorption process.
\begin{table}[hbp]
\begin{center}
\begin{tabular}{c|c|r|r}
\hline\hline
$\alpha$ & $|a_{\eta^\prime N}|$ [fm] &
 $t_{\eta^\prime p \rightarrow \eta^\prime p} [10^{-2}{\rm MeV}^{-1}]$ &
 $t_{\eta^\prime p \rightarrow \eta p}[10^{-2}{\rm MeV}^{-1}]$ \\
\hline
$-0.126 $&$ 0.075 $&$ -0.932 -0.249i$&$ 0.00843+ 0.666i $\\
$-0.193 $&$ 0.1 $&$ -1.26 -0.252i$&$ -0.0152+ 0.670i$\\
$-0.333 $&$ 0.15 $&$ -1.91 -0.261i $&$ -0.0615+ 0.678i $\\
$-0.834 $&$ 0.3 $&$ -3.85 -0.310i $&$ -0.200+ 0.699i$ \\
$-1.79 $&$ 0.5 $&$ -6.43 -0.430i $&$ -0.385+ 0.724i$ \\
$-3.93 $&$ 0.75 $&$ -9.63  -0.667i $&$ -0.615 + 0.748i $ \\
$-9.67 $&$ 1 $&$ -12.9  -1.01i $&$ -0.849 +0.767i$ \\
\hline
\end{tabular}
\end{center}
\caption{Elastic amplitude $t_{\eta^\prime p\rightarrow\eta^\prime p}$ and 
 transition amplitude $t_{\eta^\prime p\rightarrow\eta p}$ at the $\eta^\prime N$
 threshold energy calculated by the model of Ref.~\cite{etaprime}. The
 parameter $\alpha$
 indicates the coupling strength of the singlet meson to the baryons
 defined in Ref.~\cite{etaprime}, and $|a_{\eta^\prime N}|$ is the calculated
 modulus of the $\eta^\prime N$ scattering length.
\label{table:amplitude}}
\end{table}

\begin{figure}[htb]
 \includegraphics[width=0.65\linewidth]{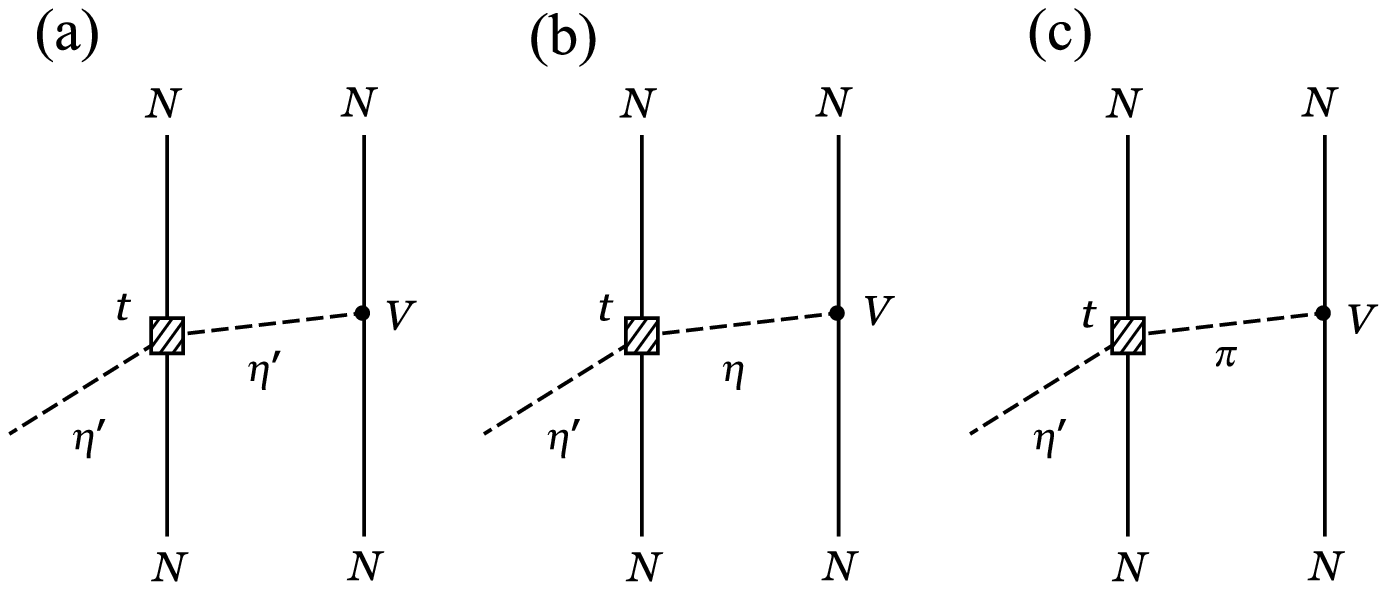}%
 \caption{Diagrams considered in the evaluation of the
  interaction of the $\eta^\prime$ meson with a pair of nucleons.  The symbol $t$ indicates the
  meson-nucleon scattering amplitude~\cite{etaprime} and $V$ the Yukawa
  vertex of each meson.
\label{fig:diagrams}}
 \end{figure}

The second order potential is evaluated using the diagrams of
Fig.~\ref{fig:diagrams}, which account for the absorption of
the $\eta^\prime$ meson by pairs of nucleons. 

\begin{figure}[htb]
 \includegraphics[width=0.5\linewidth]{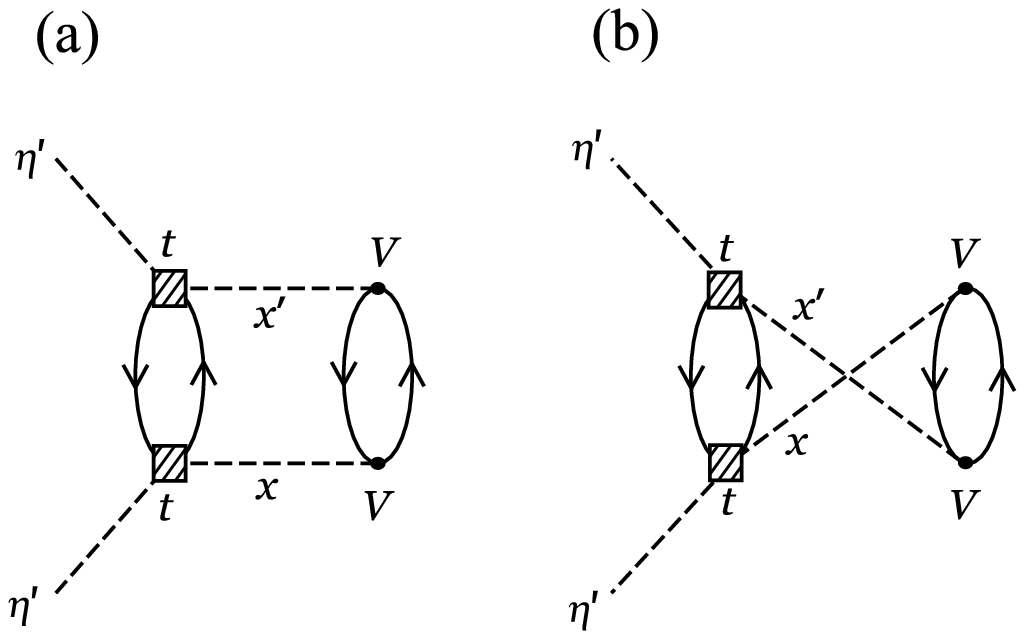}%
 \caption{Direct and crossed diagrams for the $\eta^\prime$ selfenergy in
  nuclear matter at second order. The symbols $t$ and $V$ indicate the
  scattering
  amplitude~\cite{etaprime} and the Yukawa vertex as in
  Fig.~\ref{fig:diagrams}.  $x$ and $x'$ indicate the intermediate mesons.
\label{fig:selfenergy}}
 \end{figure}

  From these open diagrams we construct the many body
 diagrams of Fig.~\ref{fig:selfenergy} for the $\eta^\prime$ selfenergy in nuclear
 matter. The imaginary part of the diagram of
 Fig. \ref{fig:selfenergy}(a), stemming from the excitation of
 intermediate two-particle two-hole ($2p2h$) states,  accounts for $\eta^\prime$ absorption by
 means of two nucleons, but there is also a
 source of imaginary part attached
 to the complex value of $t$. In addition, the diagrams also account for
 dispersive effects resulting in a real part of the $\eta^\prime$
 selfenergy. These diagrams, involving two holes, have roughly a $\rho^2$
 type behavior.

  The evaluation of the diagrams of Fig.~\ref{fig:selfenergy} requires the knowledge
 of the $t$ matrix for the $\eta^\prime  N \to \eta^\prime  N$,
 $\eta^\prime  N \to \eta  N$ and $\eta^\prime  N \to \pi  N$ transitions. Yet,
the
 $\eta^\prime  N \to \pi  N$ amplitude was found extremely small  in \cite{etaprime} , and
 experimentally the $\pi^- p \to \eta^\prime n$ cross section is known
 to be of the order of 0.1 mb \cite{Rader:1973mx}. This is substantially smaller
 than the values of order of a few mb found for $\eta^\prime  N \to \eta^\prime  N$ or
 $\eta^\prime  N \to \eta  N$ reactions  in \cite{etaprime},  in spite
 of the larger phase space involved in the  $\eta^\prime  N \to \pi  N$
 transition. 
This allows us to neglect the diagram (c) in Fig.~\ref{fig:diagrams} and consider
 only the exchange of $\eta^\prime$ and $\eta$ mesons in
 Fig.~\ref{fig:selfenergy}.

  For the Yukawa vertex we use the standard Lagrangian as,
%  second formual of pag 3
\begin{equation}
 {\cal L}_{\rm Yukawa} =
  \frac{D+F}{2f_\pi}(-\sqrt{2})\langle\bar{B}\gamma^\mu\gamma_5\del_\mu\phi
  B \rangle
+ \frac{D-F}{2f_\pi}(-\sqrt{2})\langle\bar{B}\gamma^\mu\gamma_5 B \del_\mu \phi
  \rangle \ .
\label{eq:yukawa}
\end{equation}

  The coupling of the $\eta^\prime$ is implemented adding to the
   SU(3) matrix of the pseudoscalar fields, $\phi$, the term
  ${\rm diag}(\eta_1,\eta_1,\eta_1)/\sqrt 3$.  We find the couplings for the
  octet and the singlet isospin zero pseudoscalar mesons as,
%  last two formulas of pag 4
\begin{eqnarray}
 {\cal L}^8 = \frac{1}{\sqrt{3}}\frac{D-3F}{2f_\pi}\bar{p}\gamma^\mu\gamma_5p\del_\mu\eta_8,
\label{eq:L8}\\
 {\cal L}^1 = -
  \sqrt{\frac{2}{3}}\frac{D}{f_\pi}\bar{p}\gamma^\mu\gamma_5p\del_\mu\eta_1
  \ .
\label{eq:L1}
\end{eqnarray}

  Nonrelativistically this leads to two vertex functions of the type $V
  \vec{\sigma}\cdot\vec{q}$, with $V_{\eta},  V_{\eta^\prime}$ given by
%
%  eqs page 14, 4 eqns
\begin{eqnarray}
 V_\eta = \cos\theta_P V_8 - \sin\theta_P V_1 \label{eq:Veta} \ ,\\
 V_{\eta^\prime} = \sin\theta_P V_8 + \cos\theta_P V_1 \label{eq:Veta'} \ ,
\end{eqnarray}
where $\theta_P$ is the $\eta_1$-$\eta_8$ mixing angle, for which we
take the value $14.34^{\circ}$ of Ref.~\cite{ambrosino}, and 
\begin{eqnarray}
 V_8 = \frac{1}{\sqrt{3}}\frac{3F-D}{2f_\pi} \label{eq:V8} \ ,\\
 V_1 = \sqrt{\frac{2}{3}}\frac{D}{f_\pi} \label{eq:V1} \ ,
\end{eqnarray}
with F=0465, D=0.795 \cite{boradf}. The relativistic calculation with the $\gamma_5$
  operators can be done by simply substituting $\vec{q}^{\,2}$ by the relativistic $-q^2$.

  Following standard many-body techniques, the second order
  $\eta^\prime$ selfenergy, $\Pi^{\rm 2nd}$, depicted in 
  Fig.~\ref{fig:selfenergy} is obtained from
% first  eq of page 8, remove the factor 2 of 2q2
\begin{eqnarray}
 -i\Pi^{\rm 2nd}(p_{\eta'}) &=& \sum_{x,x'=\eta,\eta^\prime}
  \int\frac{d^4q}{(2\pi)^4}iU(p_{\eta^\prime}-q)iU(q)(-it_{\eta^\prime N\rightarrow
  xN})
(-it_{x'N\rightarrow\eta^\prime N})\nonumber\\
&&\times\frac{i}{q^2-m_x^2+i\epsilon}\frac{i}{q^2-m_{x'}^2+i\epsilon}
V_x V_{x'}q^2 \ ,
\label{eq:Pi}
\end{eqnarray}
where the indices $x$ and $x'$ express the intermediate $\eta$ and/or
$\eta^\prime$ mesons and
$U(k)$ is the Lindhard function corresponding to a particle-hole ($ph$) excitation defined as,
% last eq, of page 8,
\begin{equation}
 U(k)=4\int\frac{d^3p}{(2\pi)^3}\frac{M}{E(\vec{p}\,)}\frac{M}{E(\vec{k}+\vec{p}\,)}
\left[\frac{n(\vec{p}\,)(1-n(\vec{p}+\vec{k}))}{k^0+E(\vec{p}\,)-E(\vec{p}+\vec{k})+i\epsilon}
+\frac{n(\vec{p}+\vec{k})(1-n(\vec{p}\,))}{-k^0+E(\vec{p}+\vec{k})-E(\vec{p}\,)+i\epsilon}
\right],
\label{eq:lindhard}
\end{equation}
 with the factor 4 accounting for the sum over the nucleon spin and
 isospin.  Here, the first term in the integral corresponds to the 
 direct diagram while the second term corresponds to the crossed diagram
 of the fermion loop of the $ph$ excitation.
In the present calculation, we find that the contribution of the crossed
 diagram in the first Lindhard function $U(p_{\eta^\prime}-q)$ of
 Eq.~(\ref{eq:Pi}) is small and can be neglected safely.
As for the second Lindhard function $U(q)$ of Eq.~(\ref{eq:Pi}), we
 include both direct and crossed terms in Eq.~(\ref{eq:lindhard}).
 Hence, the second order selfenergy of $\eta^\prime$ calculated here by
 Eq.~(\ref{eq:Pi}) consists of the contributions shown in
 Fig.~\ref{fig:selfenergy} (a) and (b).
%With the large momenta of the particle in the $ph$
% excitation, the Pauli blocking factor $1-n(\vec{p}+\vec{q})$ is always
% unity and can be removed.
We use average values of $\vec{p}{\,^2}$ and $\vec{p}$ to
replace
$E(\vec{p}\,)$ and $E(\vec{p}+\vec{k})$ in the denominator of
Eq.~(\ref{eq:lindhard}) as,
\begin{eqnarray}
E(\vec{p}\,)&\rightarrow &E_F=\sqrt {M_N^2+3k_F^2/5} \\
 E(\vec{p}+\vec{k}) &\rightarrow& E(\vec{k})
\end{eqnarray}
where $k_F$ is the Fermi momentum defined as,
\begin{equation}
 \frac{2k_F^3}{3 \pi ^2}= \rho(r)\ .
\end{equation}
We also use the same average value for the occupation number
$n(\vec{p}+\vec{k})$ in the numerator in Eq.~(\ref{eq:lindhard}) as,
\begin{equation}
 n(\vec{p}+\vec{k}) \rightarrow n(\vec{k}) \ .
\end{equation}
Together with these approximations, we can perform the $d^3p$ integration
by using the relation,
\begin{equation}
 4\int\frac{d^3p}{(2\pi)^3}n(\vec{p}\,) = \rho,
\end{equation}
and we finally get
\begin{eqnarray}
 -i\Pi^{\rm 2nd}(p_{\eta^\prime}) &=& - \sum_{x,x'=\eta,\eta^\prime}
  \int\frac{d^4q}{(2\pi)^4} \rho^2
\left(\frac{M}{E(\vec{q}\,)}\right)^2
(1-n(\vec{q}\,))
t_{\eta^\prime N\rightarrow xN}t_{x'N\rightarrow\eta^\prime N}V_x V_{x'}q^2
\nonumber\\
&&\times \frac{1}{p_{\eta^\prime}^0-q^0+E_F-E(\vec{q}\,)+i\epsilon} \left[\frac{1}{q^0+E_F-E(\vec{q}\,)+i\epsilon}
+\frac{1}{-q^0+E_F-E(\vec{q}\,)+i\epsilon}
\right]\nonumber\\
&&\times\frac{1}{q^2-m_x^2+i\epsilon}\frac{1}{q^2-m_{x'}^2+i\epsilon}\ ,
\end{eqnarray}
where  we have taken $M/E_F\sim 1$ .
Performing the $q^0$ integration using Cauchy theorem and summing up
the contributions of all poles, we obtain a simplified expression of the $\eta^\prime$
self-energy
$\Pi^{\rm 2nd}(p_{\eta^\prime})$
which is calculated numerically.  The second order optical potential
$V^{\rm 2nd}_{\rm opt}$ is obtained as
$
 V^{\rm 2nd}_{\rm opt} = {\Pi^{\rm 2nd}}/{2\omega_{\eta^\prime}} \ .
$

%eq. last of page 10, beginning of page 11.

  In order to obtain the optical potential in finite nuclei we use the
  local density approximation, substituting $\rho$ by $\rho(r)$, where
  $\rho(r)$ is 
{assumed to be an empirical Woods-Saxon form. }
This
  prescription was proved to be exact for $s$-wave \cite{juanito}, which
  is the only one we consider here.

\section{Results}

In Table~\ref{tab:potential}, we show the numerical results of the
optical potential $V_{\rm opt}$
at normal nuclear matter density
$\rho=\rho_0$ for
different values of the scattering length $a_{\eta^\prime N}$ corresponding to
those of Table~\ref{table:amplitude}.
We find that the strength of the optical potential at normal nuclear
density is $V_{\rm opt}|_{\rho_0} = -(8.71+1.82i)$ MeV for the case
$|a_{\eta^\prime N}|=0.1$ fm, which is consistent to the data in
Ref.~\cite{Moskal:2000pu}.  In the case with $|a_{\eta^\prime N}|=0.5$~fm, we have
$V_{\rm opt}|_{\rho_0}=-(45.14+5.47i)$~MeV, giving rise to a
width $\Gamma=-2\,{\rm Im}V_{\rm opt}\sim 10$~MeV , which could be a
reasonable extrapolation to low $\eta^\prime$ momentum values of the
transparency ratio data of Ref.~\cite{nanova}. 
If we consider an extreme
case with $|a_{\eta^\prime N}|=1$~fm, we obtain a deep potential with a relatively
small imaginary part, $V_{\rm opt}|_{\rho_0}=-(91.81+17.21i)$ MeV.
The potential in this extreme case has  similar features as the one
discussed in Refs.~\cite{Nagahiro:2006dr,Jido:2011pq}; however, the
corresponding scattering length is much larger than
the experimental value reported in Ref.~\cite{Moskal:2000pu}.

%In table 1 put also the values of $V_{opt}$ at rho0 for comparison with this table.
\begin{table}[htbp]
\begin{center}
\begin{tabular}{c||c|c||cc|cc}
& \multicolumn{2}{c||}{potential depth at $\rho=0.17$ fm$^{-3}$}
& \multicolumn{4}{c}{bound states :
 $n\ell$ (B.E., $\Gamma$) [MeV] }
 \\\hline
$|a_{\eta^\prime N}|$ [fm] & {$V_{\rm opt}^{\rm 1st}$ [MeV]  } & {$V_{\rm
	 opt}^{\rm 2nd}$ [MeV] }
& \multicolumn{2}{c|}{$^{11}$C} & \multicolumn{2}{c}{$^{39}$Ca}
\\\hline\hline
$ 0.075 $&$ -6.36  -1.70i $&$ -0.07  -0.05i $& - & - & - & - \\\hline
$ 0.1 $&$ -8.61  -1.72i $&$ -0.10  -0.10i $& - & - &$ 0s $&$ (-1.61, 2.01)$ \\\hline
$ 0.15 $&$ -13.04  -1.78i $&$ -0.19  -0.23i $& - & - &$ 0s $&$ (-4.51, 2.78)$ \\\hline
$ 0.3 $&$ -26.26  -2.11i $&$ -0.56  -0.91i $&$ 0s $&$ (-4.65, 2.55) $&$ 0s $&$ (-15.33, 5.09)$ \\
       &  &  &  &  &$ 0p $&$ (-5.43, 3.74)$ \\\hline
$ 0.5 $&$ -43.83  -2.93i $&$ -1.31 -2.54i $&$ 0s $&$ (-14.50, 5.95) $&$ 0s $&$ (-31.79, 10.05)$ \\
     &  &  &  &    &$ 1s $&$ (-4.29, 5.16)$ \\
     &  &  &  &    &$ 0p $&$ (-18.94, 8.38)$ \\\hline
$ 0.75 $&$ -65.62 -4.55i $&$ -2.57  -5.73i $&$ 0s $&$ (-29.34, 12.80) $&$ 0s $&$ (-53.91, 20.14)$ \\
       &  &  &$ 0p $&$ (-7.22, 7.52) $&$ 1s $&$ (-18.70,13.70)$ \\
       &  &  &  &  &$ 0p $&$ (-38.44, 17.60)$ \\
       &  &  &  &  &$ 1p $&$ (-3.06, 8.61)$ \\\hline
$ 1 $&$ -87.72  -6.86i $&$ -4.09  -10.35i $&$ 0s $&$ (-46.10, 23.34) $&$ 0s $&$ (-77.65, 35.42)$ \\
     &    &  &$ 1s $&$ (-0.29, 4.99) $&$ 1s $&$ (-36.58, 26.34)$ \\
     &    &  &$ 0p $&$ (-18.73, 15.78) $&$ 0p $&$ ( -59.96, 31.66)$ \\
     &    &  &  &  &$ 1p $&$  ( -16.33, 20.56)$ \\\hline
\end{tabular}
\end{center}
\caption{Numerical results of the optical potential strength at normal
 nuclear density and results of the $\eta^\prime$-nucleus bound states in
 $^{11}$C and $^{39}$Ca
 are shown for various $|a_{\eta^\prime N}|$ values.
 The strengths of the lowest order $(V_{\rm opt}^{\rm 1st})$ and the
 second order $(V_{\rm opt}^{\rm 2nd})$ potentials are
 listed separately.
}
\label{tab:potential}
\end{table}

As we can see, for values of $|a_{\eta^\prime N}|$ of the order of 0.1
fm, the second 
order potential is smaller than the first one, but for values of
$|a_{\eta^\prime N}|$ of the 
order of 0.5 fm the imaginary part of the second order potential
acquires a value comparable in size to that of the first order
contribution. If we take a value of $|a_{\eta^\prime N}|$ of the order
of 1 fm, the 
imaginary part of the second order potential is 50\% larger than the one
of lowest order, which indicates the breakdown of the low-density
expansion and that absorption by more nucleons may start being
sizable. Since the 
imaginary part from additional multinucleon absorption processes not
considered here will always be negative, 
we expect the imaginary part of the self-energy to be larger than the present
results for the $|a_{\eta^\prime N}|\gtrsim 0.5$ fm cases.

   We solve the Klein-Gordon equation and look for bound states with
$V_{\rm opt}$ for different nuclei. The results are shown in
Table ~\ref{tab:potential} for $^{11}$C and $^{39}$Ca.  We find
$\eta^\prime$ bound states in $^{11}$C for $|a_{\eta^\prime N}|\gtrsim 0.3$~fm and
in $^{39}$Ca for $|a_{\eta^\prime N}|\gtrsim 0.1$ fm.  The level spacings of
the bound states are smaller than the widths for some cases as
shown in Table~\ref{tab:potential}.  Thus, there would be a possibility
to observe these states experimentally.

\begin{figure}[htb]
 \includegraphics[width=0.5\linewidth]{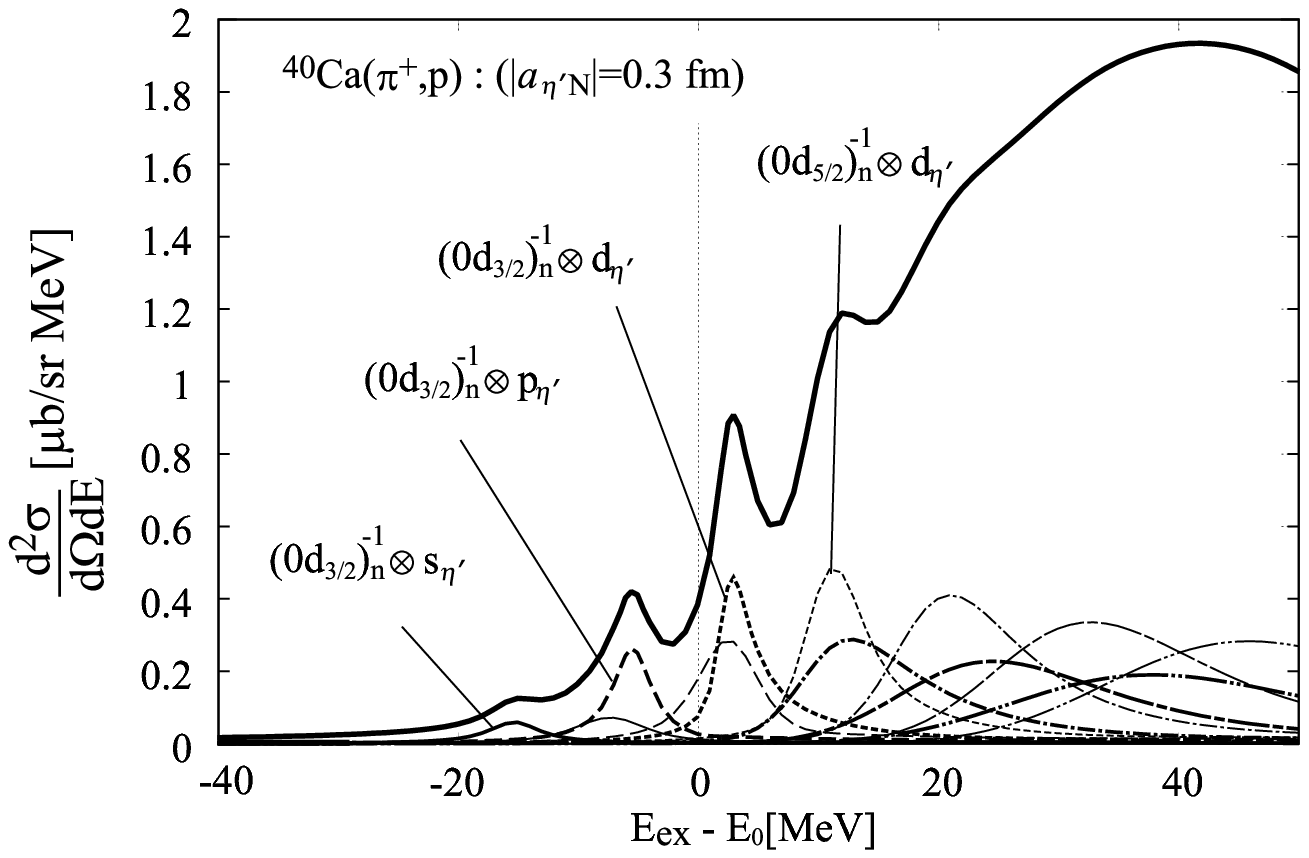}%
 \caption{Calculated spectra of $\eta^\prime(958)$ mesic nuclei formation in
  the $(\pi^+,p)$ reaction on a $^{40}$Ca target at $p_\pi=1.8$ GeV/c for the
 case
  $|a_{\eta^\prime N}|=0.3$ fm, as functions of the excitation energy
 $E_{\rm ex}-E_0$, where $E_0$ is the $\eta^\prime$ 
  production threshold energy.  The total spectrum is shown by the
  thick solid line, and the dominant [n-hole$\otimes\eta^\prime$]
 configurations are also shown in the 
  figures.  The neutron-hole states are indicated as $(n\ell_j)_n^{-1}$
  and the $\eta^\prime$ states as $\ell_{\eta^\prime}$.
  The elementary cross section is estimated to be 100 $\mu$b/sr  in
  the laboratory frame~\cite{Rader:1973mx,Nagahiro:2010zz}.
\label{fig:spectra40Ca}
}
 \end{figure}

\begin{figure}[htb]
 \includegraphics[width=0.5\linewidth]{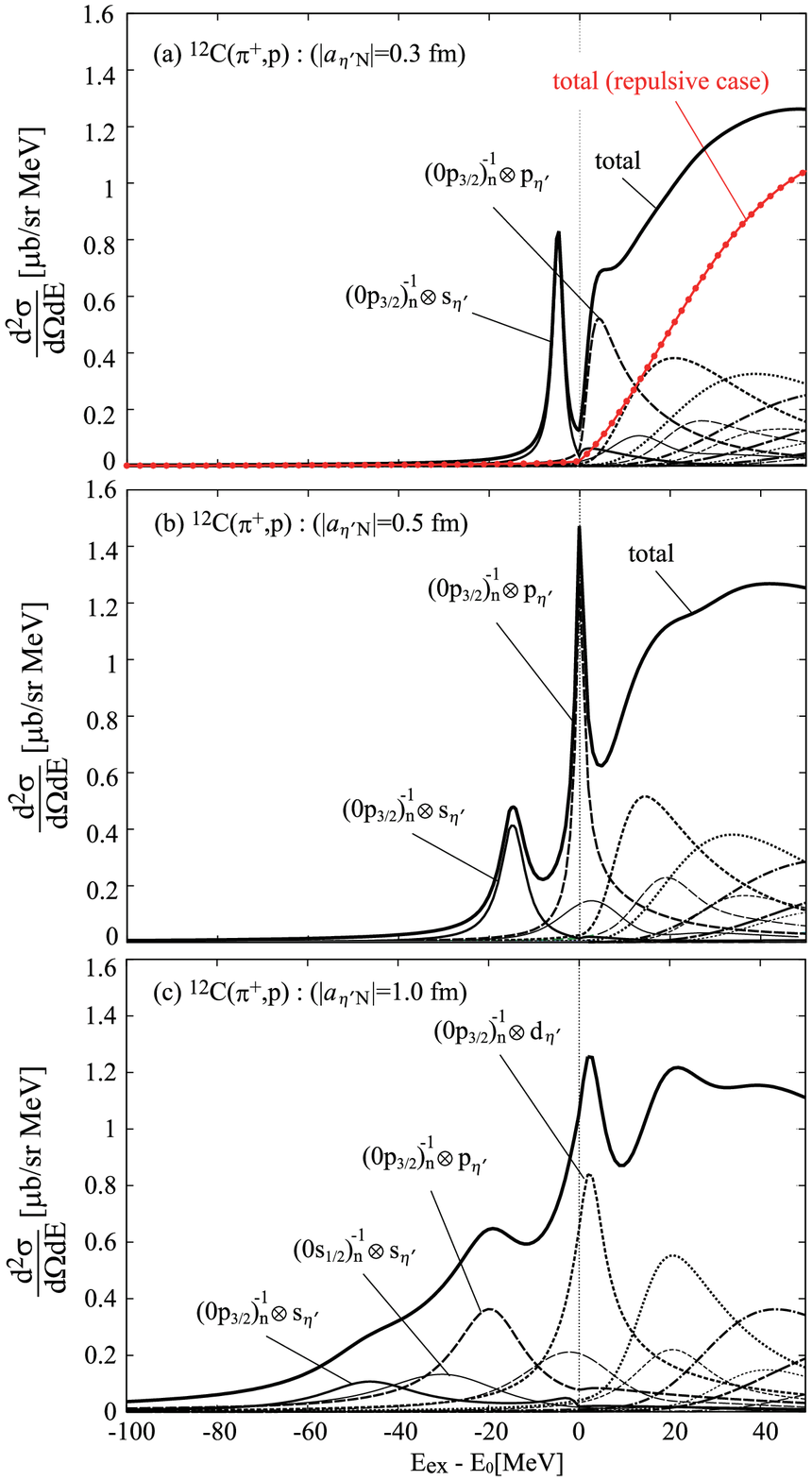}%
 \caption{(color online) Calculated spectra of $\eta^\prime(958)$ mesic nuclei
  formation in
  the $(\pi^+,p)$ reaction on a $^{12}$C target at $p_\pi=1.8$ GeV/c for the
  (a)
  $|a_{\eta^\prime N}|=0.3$ fm, (b) $|a_{\eta^\prime N}|=0.5$
  fm, and (c)  $|a_{\eta^\prime N}|=1.0$ fm cases,
as functions of the excitation energy $E_{\rm ex}-E_0$, where $E_0$ is the $\eta^\prime$
  production threshold energy.
The potential is assumed to be attractive as explained in the text.
The total spectra are shown by the
  thick solid lines, and the dominant [n-hole$\otimes\eta^\prime$]
  configurations are also shown in the 
  figures.
The total spectrum for the repulsive potential case is also shown for
  $|a_{\eta^\prime N}|=0.3$ fm in panel (a) by the solid circles with
  the thin line.
 The neutron-hole states are indicated as $(n\ell_j)_n^{-1}$
  and the $\eta^\prime$ states as $\ell_{\eta^\prime}$.
  The elementary cross section is estimated to be 100 $\mu$b/sr  in
  the laboratory frame~\cite{Rader:1973mx,Nagahiro:2010zz}.
\label{fig:spectra}
}
 \end{figure}

As an example, {by using the the same
theoretical model of Refs.~\cite{Nagahiro:2008rj,Nagahiro:2010zz},}
we calculate theoretically the expected formation cross
section of the $\eta^\prime$ bound states in the $(\pi^+,p)$ reaction
at {the pion beam momentum} $p_\pi=1.8$~GeV/c, {which can be available
at the J-PARC facility.}
{We consider the forward reactions where the emitted proton is
observed at 0 degrees in the laboratory frame to reduce the momentum
transfer.  The momentum transfer, however, is larger than 200 MeV$/c$ even
at the forward angle and cannot be zero at any kinematics in
this reaction.}
{The cross section for this reaction is evaluated using the Green's
function method~\cite{Morimatsu:1985pf} {in which we solve the Klein-Gordon equation
both for the bound and unbound $\eta'$ meson with the $\eta'$ optical
potential.}  The pion and proton wave 
functions are also 
distorted using the eikonal approximation. 
The neutron in the target nucleus is assumed to be described by a
simple
harmonic oscillator wavefunction.}
{The energy of the emitted proton determines the energy of the
$\eta'$-nucleus system uniquely.}
We show the 
calculated spectra in Figs.~\ref{fig:spectra40Ca} and \ref{fig:spectra}
{as functions of the excitation energy $E_{\rm
ex}-E_0$ defined as
\begin{equation}
 E_{\rm ex} - E_0 = - B_{\eta'} + [S_n(j_n)-S_n({\rm ground})],
\end{equation}
where $B_{\eta'}$ is the $\eta'$ binding energy and $S_n(j_n)$ the
neutron separation energy from the neutron single-particle level
$j_n$.  $S_n({\rm ground})$ indicates the separation energy from the
neutron level corresponding to the ground state of the daughter
nucleus.  $E_0$ is the $\eta'$ production threshold energy.}
In  Figs.~\ref{fig:spectra40Ca},
\ref{fig:spectra}(a) and \ref{fig:spectra}(b), we find that a clear
peak structure appears .  The result with 
the deep potential shown in Fig.~\ref{fig:spectra}(c) has similar features as
those found in Ref.~\cite{Nagahiro:2010zz}.
We also show the spectrum for the repulsive potential case in
Fig.~\ref{fig:spectra}(a), where we assume an opposite (repulsive)
sign for Re($V_{\rm opt}$) in the $|a_{\eta^\prime N}|=0.3$~fm case.
The spectra for  nuclei
heavier than $^{39}$Ca have an overlap of different [n-hole$\otimes\eta^\prime$]
configurations that smear out the individual peaks
and would not be suited for experimental searches.

  The results shown in Figs.~\ref{fig:spectra40Ca} and \ref{fig:spectra}
  deserve some
  comments. The peaks appearing in the most bound region correspond to a
  situation where the neutron has been removed from
  the less bound orbit, namely
{$(0d_{3/2})_n$ for the $^{40}$Ca target
  (Fig.~\ref{fig:spectra40Ca}) and
$(0p_{3/2})_n$ for the $^{12}$C target (Fig.~\ref{fig:spectra})},
  and the $\eta^\prime$ is in the most bound $0s$ state. The final
  nucleus is thus left in its ground state which is stable.
{We also take into account}
situations where the neutron has been taken from an inner
  orbit, leaving a hole in the final nucleus which is
  then in a particular excited state. These states decay and their width
  is tied to the imaginary part of the nucleon nucleus potential for
  energies below the Fermi energy
  \cite{reportmahaux,pedrosemi,pandha}.
%
%ANGELS HELP !!!!. This width
%  can be of the order of 10-15 MeV and should be added to the peaks
%  obtained ????????, which would dilute the peaks appearing in the
%  unbound region and some of those appearing in the bound region in the
%  case of an optical potential with large strength ( see
%  Fig.~\ref{fig:spectra}, lower figure).  In the upper and lower figures
%  of Fig.~\ref{fig:spectra} we can see that the removal of the outer
%  nucleon and production of the $\eta^\prime$ in a d-wave state also has a
%  peak in the unbound region. This is due to ........
%
The widths of the hole states are taken into account in the present
calculation.  The width of the neutron-hole states in $^{39}$Ca have
been estimated to be $\Gamma=7.7$ MeV $((1s_{1/2})^{-1})$, 3.7 MeV
$((0d_{5/2})^{-1})$, 21.6 MeV $((0p_{3/2,1/2})^{-1})$, and 30.6 MeV
$((0s_{1/2})^{-1})$ from the data in Ref.~\cite{Nakamura:1974zz},
considering the width of the ground state $(0d_{3/2})^{-1}$ to be 0 and
assuming the same widths for neutron-hole states as those of proton
holes.  As for $^{11}$C, we have used the data in Ref.~\cite{11Chole}
and the widths are $\Gamma((0s_{1/2})^{-1})=12.1$ MeV for the
excited state and
$\Gamma((0p_{3/2})^{-1})=0$ MeV for the ground state.
The first and second
peaks from the left
in the total spectrum for the $^{40}$Ca target, shown in
Fig.~\ref{fig:spectra40Ca}, correspond to cases in which the neutron is
removed from the outermost single-particle orbit ($(0d_{3/2})_n$) and
the $\eta^\prime$ is left bound in the daughter nucleus $^{39}$Ca in the
lower $s$ and $p$ states, respectively.  The third peak just above the
threshold 
($E_{\rm ex}-E_0=0$) is dominated by an unbound $d$-wave component of
the $\eta^\prime$ accompanied by 
the removal of the outer neutron.
{While} there are no bound
$d$ states of $\eta^\prime$ in this case, {as can be seen in the results
shown in Table~\ref{tab:potential},
the attractive $\eta^\prime$-nucleus interaction pulls this
low energy scattering wave of the $\eta^\prime$ closer to the daughter
nucleus enhancing its overlap 
with the nucleon wavefunctions and consequently producing a
larger cross section.}
This is the so-called threshold enhancement of the quasi-elastic
(unbound) $\eta^\prime$ contributions around the production threshold.
{Therefore, we can consider} this
enhancement to give an
indication of the attractive $\eta^\prime$-nucleus interaction if it is
observed. 
{The fourth peak from the left at $E_{\rm ex}-E_0 \sim 12$ MeV has
the same origin as the third peak of the threshold enhancement, except
that it is accompanied by the removal of the inner neutron state $(0d_{5/2})_n$.
}
{Extra peaks in the higher energy region have the same interpretation but
correspond to the higher angular momentum partial waves.}

We can observe
  that in an intermediate mass nucleus like $^{40}$Ca, the
  capture of the $\eta^\prime$ in the most bound $(0s)$ state with removal of
  an outer nucleon, produces a peak with too small strength, diluted
  with the tail from other states, which would not be suited for
 experimental observation. The production of a $0p$ $\eta^\prime$ state, however,
  leads to a distinct signal. In the case of a lighter nucleus like
  $^{12}$C, a peak in the bound region can be observed, as long as
$|a_{\eta^\prime N}|$ does not differ too much from the value where it
  starts producing bound states (see the upper and middle panels of
  Fig.~\ref{fig:spectra}). 
Curiously, it does not help to have a larger strength for the optical
  potential (lower panel in Fig.~\ref{fig:spectra}).
Indeed, the peaks corresponding to the production of the $\eta^\prime$ in the most bound
  state have small strength and are diluted due to the overlap of
  competing contributions from the removal of nucleon states from inner and wider
  orbits. Only a feeble signal over a large background stands up in the
  bound region corresponding to the 
  production of the  $\eta^\prime$ in the $0p$ bound state with the removal of the
  outer nucleon.

An interesting observation of the numerical results for attractive
potentials is the robustness of
the appearance of the peak structures in the spectra around the meson production
threshold, in contrast to the spectrum calculated with the repulsive
potential, where we only find a smooth quasifree contribution without any
peak structure, as shown in Fig.~\ref{fig:spectra}(a).
The origin of the structures depend on the target nucleus and
the $\eta^\prime$-nucleus potential strength, and they can be associated
to genuine $\eta^\prime$ bound states and/or 
to an enhancement of the quasifree contribution.  In any case, they give
an indication of the attractive feature of 
the potential and can provide new information on the properties of the
$\eta^\prime$ meson in nuclei. 
The robustness of the appearance of structures for an attractive
interaction, which is independent on the details of the theory, could be
helpful for actual experiments even if subsequent more refined measurements might be
required to properly identify the origin of the peak structure in the spectrum.

 In summary, we have observed that, depending on the strength
  of the potential, some nuclei are better suited than others to
  eventually find peaks that can be identified with the production of
  possible bound $\eta^\prime$ states in nuclei. The calculated spectra
  shown in Figs.~\ref{fig:spectra40Ca} and \ref{fig:spectra} also
  indicate the 
  experimental resolution needed to observe those peaks.  We note that
  the relatively small width of these states has made the appearance 
  of clean peaks possible, in contrast with the situation with $\bar{K}$ mesons
  in nuclei where the widths are very large
  \cite{angelskmed,koch,okumura}.

\section{Conclusions}

We have done a calculation of the $\eta^\prime$-nucleus optical potential,
starting from the lowest order term in the nuclear density and adding
the second order term which accounts for $\eta^\prime$ absorption by pairs of
nucleons. The strength of the potential is unknown, because of the
limitations of the theory, where one has a free parameter tied to the
coupling of the singlet of mesons to the baryons, for which the only
constraint is the poorly known $\eta^\prime N$ scattering length. At present, we
can not even predict the sign, so we cannot claim the existence of bound
$\eta^\prime$ states in nuclei. However, the free parameter of the theory is
related to the $\eta^\prime N$ scattering length and, assuming different
values for it, we have obtained $\eta^\prime$-nucleus potentials with 
varying strength that produced bound states in several nuclei.
The potential strength can be made similar to that expected in
Refs.~\cite{Nagahiro:2006dr,Jido:2011pq}, but it requires a very large
scattering length value of $|a_{\eta^\prime N}|\simeq 1$ fm in the present calculation.
One welcome
feature of the theory is that, within a wide range of values of the
scattering length, the imaginary part of the potential is reasonably
smaller than the real part and this leads to bound  $\eta^\prime$ states with
a width smaller than the separation between the levels. In such cases,
one may in principle expect to see distinguishable peaks in some experiments.

As an example of a possible experiment we have presented results for the
$(\pi^+,p)$ reaction, which can be measured at J-PARC, complementary to
the reaction investigated in \cite{marianamaybe} at ELSA.
{We note that $(\pi^-,n)$ reactions  on isospin symmetric targets as
the ones used here,  $^{12}$C and $^{40}$Ca, would have given similar
$\eta^\prime$ production spectra. Given the fact that there is a
proposal at J-PARC that plans to use this reaction to look for $\omega$
bound states \cite{ozawa} measuring forward going neutrons, the results
that we get would be useful if this reaction is also done to look for
$\eta'$ states.  
}
{In the photoproduction case, we can naively expect to obtain
similar spectral shapes as shown in this paper with a
smaller magnitude of the formation cross section, of order of 5--10
nb/sr/MeV~\cite{Nagahiro:2004qz,Nagahiro:2006dr}.} 

We found that, for certain values of
the scattering length and some nuclei, clean peaks could be resolved.
The bound state peaks are visible in light or medium nuclei and
correspond to situations where the potential is not 
too deep. The peaks appear at binding energies
smaller than 20 MeV and their widths are of the order of 10 MeV. This
certainly requires having a good experimental resolution.
It is also interesting to observe that the structure of the spectrum around
threshold gives clear indications about the character (attractive or
repulsive) of the $\eta^\prime$-nucleus interaction. 
 In the present
situation, where not even the sign of the scattering length, and hence of
the $\eta^\prime$-nucleus potential, is known, an experiment searching for bound
$\eta^\prime$ in nuclei, either producing positive or negative results, would
provide a welcome information on the properties of the elementary
$\eta^\prime N$ interaction, 
putting constraints on the size of the scattering length and eventually 
determining its unknown sign.

\section*{Acknowledgments}
{We appreciate the useful comments by V.~Metag and
M.~Nanova. }
This work is partly supported by  projects FIS2006-03438, FIS2008-01661
 from the Ministerio de Ciencia e Innovaci\'on (Spain), by the
 Generalitat Valenciana in the program Prometeo and
 by the Ge\-ne\-ra\-li\-tat de Catalunya contract 2009SGR-1289.
This work is also supported by the Grant-in-Aid for Scientific Reserach
 (No.~22105510 and No.~20540273) in Japan.
 This research is part of the European
 Community-Research Infrastructure Integrating Activity ``Study of
 Strongly Interacting Matter'' (acronym HadronPhysics2, Grant
 Agreement n. 227431) and of the EU Human Resources and Mobility
 Activity ``FLAVIAnet'' (contract number MRTN--CT--2006--035482),
 under the Seventh Framework Programme of EU.

\end{document}